# STAR FORMATION IN NORMAL AND BARRED CLUSTER SPIRALS

C. MOSS
Institute of Astronomy
Madingley Road, Cambridge CB3 0HA, U.K.

M. WHITTLE
Department of Astronomy, University of Virginia
Charlottesville, VA 22903, U.S.A.

J.E. PESCE
Space Telescope Science Institute
3700 San Martin Drive, Baltimore, MD 21218, U.S.A.

H. NAVARRO
Royal Greenwich Observatory
Madingley Road, Cambridge CB3 0EZ, U.K.

**Abstract** – An objective prism H$\alpha$ survey has shown that there is a population of early type spiral galaxies in nearby clusters with strong central bursts of star formation which could be due to galaxy–galaxy tidal interactions. Such galaxies are rarely found in the field.

## 1. INTRODUCTION

An objective prism H$\alpha$ survey of eight nearby clusters has been undertaken to compare star formation rates in cluster and field spirals (see [1] and [2]). A total of 232 CGCG spiral galaxies were surveyed, of which 42% were detected in emission.

Figure 1: Intensity contours of H$\alpha$ emission from CCD observations for CGCG 126.104 (left) and CGCG 127.049 (right). The outermost contours for each galaxy have a maximum diameter of approximately 30 arcmin and roughly coincide with the outer boundary of the optical disk. For CGCG 126.104 the H$\alpha$ emission is classified on the prism plates as *diffuse* and appears to come from HII regions distributed in the disk of the galaxy, while for CGCG 127.049 the emission is classified as *compact* and is strongly concentrated in the central region of the galaxy.

The spectra on the prism plate consist of a two-dimensional image of the H$\alpha$ emission, superposed on the continuum spectrum. Galaxy emission was classified as *compact* or *diffuse* according to its appearance on the prism plates. *Compact* emission is much brighter than the underlying continuum and sharply delineated from it. It is almost always centred on the nucleus of the galaxy in a small region (median diameter $7'' = 2.0h^{-1}$ kpc). *Diffuse* emission is only slightly brighter than the continuum and has an indistinct appearance spanning a larger region (median diameter $18'' = 5.2h^{-1}$ kpc). The *compact* emission is considered to result from a burst of star formation at the centre of the galaxy, whereas *diffuse* emission results from more normal processes of star formation in the spiral disk. This interpretation is supported by higher resolution CCD observations of a small subset of the emission-line galaxies (see Figure 1).

## 2. SURVEY RESULTS

The emission-line and non emission-line spirals have similar distributions in absolute magnitude. Barred and unbarred spirals are equally likely to have emission. However barred galaxies are more likely to have *compact* emission than unbarred spirals (significant at the $3\sigma$ level). Galaxies with a disturbed morphology are more likely to have *compact* emission than undisturbed galaxies (significant at the $7\sigma$ level). However disturbed and undisturbed galaxies are equally likely to have *diffuse* emission. Moreover, disturbed spirals are more likely to have a nearby companion than undisturbed spirals (significant at the $6\sigma$ level), and unbarred spirals showing *compact* emission are also more likely to have a nearby companion (significant at the $3.7\sigma$ level). These results can be understood if the *compact* emission is due a central burst of star formation triggered either by the barred structure of the galaxy or by galaxy–galaxy tidal interactions. The *diffuse* emission is likely to be more normal star formation in the disk of the galaxy (see [2]).

A comparison between H$\alpha$ emission in magnitude-limited samples of field spirals (from [3]) and spirals in the three clusters Abell 347, 1367 and 1656, shows that whereas late-type cluster spirals (Sc, Sc–Irr) are *less* likely to show emission than spirals of the same type in the field (significant at the $4.5\sigma$ level), early-type cluster spirals (Sa, Sab) are *more* likely to show emission than spirals of the same type in the field (significant at the $3.3\sigma$ level). Indeed there is a sub-sample of early-type cluster spirals with strong H$\alpha$ emission ($W_\lambda \geq 20$ Å) which have no counterparts in the field. The emission from these galaxies is predominantly *compact* emission, and is likely to be due to tidally-induced star formation from galaxy–galaxy interactions. There is increasing evidence that such interaction-induced star formation was more common in clusters at earlier epochs (e.g. [4]).

**Acknowledgements** – We would like to thank Reynier Peletier for assistance with CCD observations.

# References

[1] Moss, C., Whittle M. & Irwin, M.J. 1988. *Mon. Not. R. astr. Soc.*, **232**, 381.

[2] Moss, C., & Whittle, M. 1993. *Astrophys. J.*, **407**, L17.

[3] Kennicutt, R.C. & Kent, S.M. 1983. *A.J.*, **88**, 1094.

[4] Couch, W.J., Ellis, R.S., Sharples, R.M. & Smail, I. 1993. *ASP Conference Ser.*, **51**, eds. G. Chincarini, A. Iovino, T. Maccacaro & D. Maccagni, p. 240.